%% file: main.tex
\documentclass[a4paper,11pt]{article}
\usepackage{jinstpub} % for details on the use of the package, please see the JINST-author-manual
\ifdefined\pdfminorversion\pdfminorversion=7\relax\fi
\title{\boldmath Studies of dual ions arriving time reconstruction in N$\nu$DEx time projection chamber}

\author[a]{Tianyu Liang}
\author[b,c]{Hulin Wang}
\author[b,c]{Kai Chen}
\author[b,c]{Dongliang Zhang}
\author[b,c]{Chaosong Gao}
\author[b,c]{Xiangming Sun}
\author[b,c]{Feng Liu}
\author[b,c]{Jun Liu}
\author[a]{Hui Zhang}
\author[d]{Chengui Lu}
\author[d]{Yichen Yang}
\author[d]{Chengxin Zhao}
\author[d]{Hao Qiu}

\affiliation[a]{Institute of High Energy Physics, Chinese Academy of Science,\\19,Yuquan Rd. BeiJing, China}

\affiliation[b]{PLAC, Key Laboratory of Quark \& Lepton Physics (MOE), Central China Normal University,\\152, Luoyu Rd. Wuhan, China}
\affiliation[c]{Hubei Provincial Engineering Research Center of Silicon Pixel Chip \& Detection Technology,\\152, Luoyu Rd. Wuhan, China}
\affiliation[d]{Institute of Modern Physics, Chinese Academy of Sciences,\\509, Nanchang Rd., Lanzhou, China}

\emailAdd{liangty@ihep.ac.cn}

\abstract{
The N$\nu$DEx experiment aims to search for neutrinoless double-beta decay of \(^{82}\)Se with a high-pressure \(^{82}\)SeF\(_6\) negative-ion time projection chamber. In the assumed two-carrier model, the arrival-time separation of SeF\(_5^-\) and SeF\(_6^-\) can estimate the longitudinal coordinate without an external trigger. We compare FFT deconvolution and direct waveform fitting on simulated Topmetal-S charge-sensitive-amplifier waveforms.

With the fixed amplifier response, direct fitting and FFT accepted 873 and 855 of 873 geometry-selected single depositions, respectively. On the 855 common events, the arrival-time separation RMSEs were 0.180 and 0.222 ms. Numerical depth calibration gave mean \(z\) residuals of 0.036 and 0.075 cm and RMSEs of 0.192 and 0.208 cm on 848 common events. With sampled CSA shaping, FFT accepted the same 855 waveforms and direct fitting accepted 866 of 873. On the four-way common sample of 855 events, the direct-fit separation RMSE was 0.180~ms with fixed shaping and 0.217~ms with sampled shaping. An accepted near-readout direct fit outside the FFT sample had a 184~ms separation error, giving an RMSE of 6.25~ms over all accepted sampled-response direct fits.

With the fixed response, the full-track fit retained 1,185 of 1,810 observed pads in 100 selected Geant4 events. The median event-level depth RMSE was 0.184 cm. Subtracting each event's mean Monte Carlo residual gave a median event-centred RMS of 0.102 cm as a diagnostic of relative track shape. The absolute depth remains sensitive to the common event time. Validation with detector data is needed to assess the effects of diffusion, electronics noise, and inter-pad response.
}


\begin{document}
\maketitle
\flushbottom

\section{Introduction}
\label{sec:intro}

Time projection chambers (TPCs) have been widely employed in searches for neutrinoless double-beta decay ($0\nu\beta\beta$), including the EXO~\cite{MAuger_2012}, nEXO~\cite{Adhikari_2022}, NEXT~\cite{next2023}, and PandaX-III~\cite{PandaX2017} experiments. The N$\nu$DEx experiment~\cite{nvdexcdr2023} proposes a high-pressure ion TPC containing approximately 100~kg of SeF$_6$ gas to search for the $0\nu\beta\beta$ decay of $^{82}$Se.

\begin{figure}
    \centering
    \includegraphics[width=0.8\linewidth]{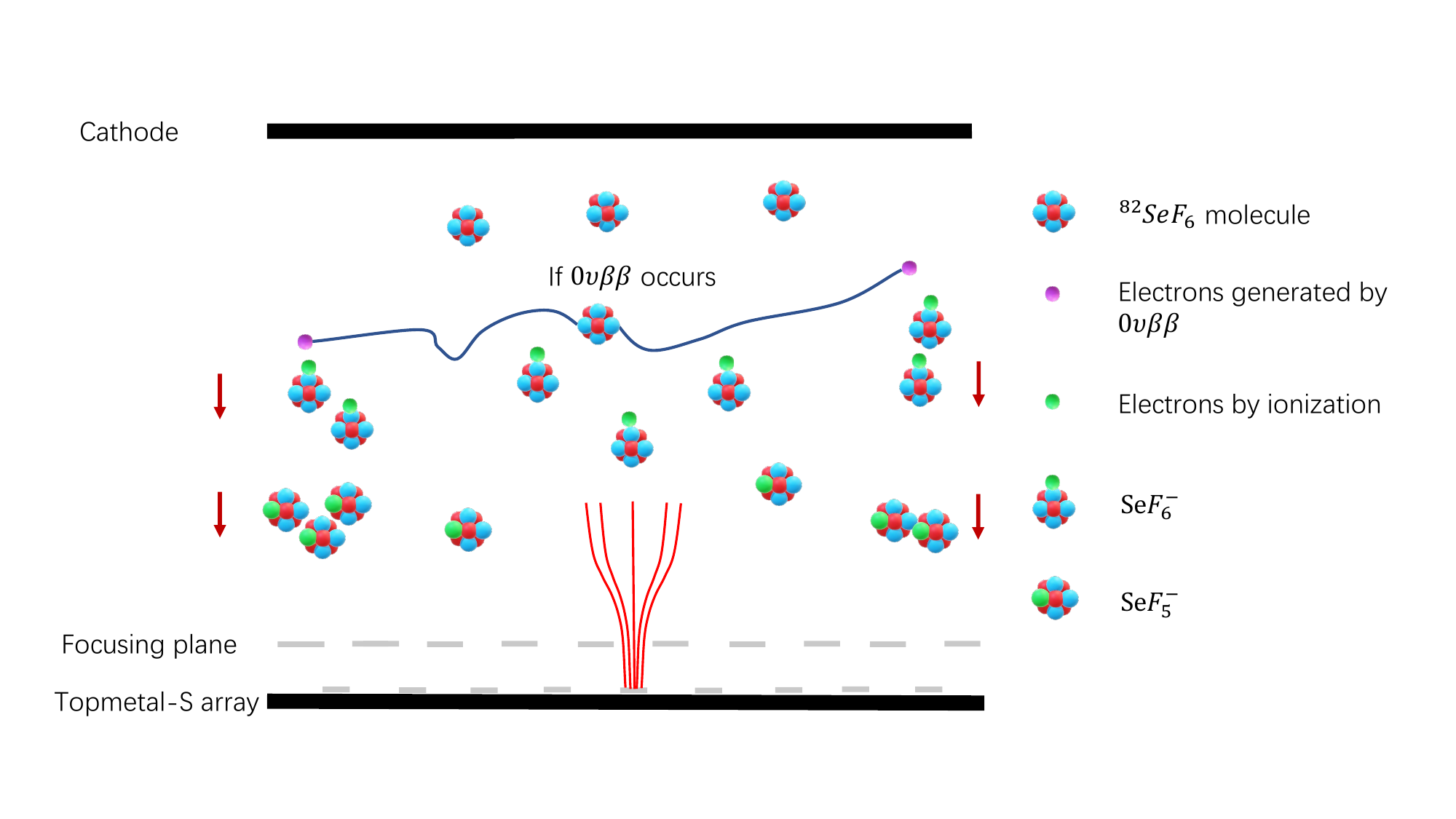}
    \caption{The conception diagram of N$\nu$DEx TPC. The focusing hole could not only guide the charged particle but also help to improve the amplitude of signal.}
    \label{fig:nvdexdiag}
\end{figure}

Different with conventional electron-drift TPCs, electrons generated by ionization in SeF$_6$ are captured through electron attachment, resulting in negative-ion transport. As is shown in Fig~\ref{fig:nvdexdiag}, under the assumption of two kinds of ions, SeF$_5^-$ and SeF$_6^-$ would be formed from charged particle ionization in SeF$_6$~\cite{Nygren_2018}, the arriving time difference ($\Delta t$) and velocities ($v_{SeF_5^-} ,v_{SeF_6^-}$) between SeF$_5^-$ and SeF$_6^-$ could be used for z coordination reconstruction without external trigger system.

For a uniform electric field, the arrival-time separation $\Delta t=t_6-t_5$ is related to the initial $z$ coordinate by
\begin{equation}
    z=z_0+\frac{v_5v_6}{v_5-v_6}\,\Delta t,
    \label{eq:reco}
\end{equation}
where $v_5>v_6>0$ are the drift-speed magnitudes of SeF$_5^-$ and SeF$_6^-$, respectively. Throughout this work, $z$ is the absolute Garfield++ coordinate, increasing away from the readout. The readout reference plane is at $z_0=0.051$~cm, so the axial drift distance in this approximation is $L=z-z_0$.

Another important feature of the N$\nu$DEx TPC is the direct collection of negative ion charge carriers without avalanche. The Topmetal-S chip~\cite{Liang_2024} has been developed for direct charge readout and is intended for use in the N$\nu$DEx TPC readout plane. A simulation framework incorporating the properties of SeF$_6$, negative-ion transport, as well as the electronic response of Topmetal-S has been developed~\cite{Liang:2025vkv}.

Although the arriving time difference provides a means of z coordination reconstruction, extracting this information directly from the readout waveform is not straightforward. The induced currents from the two ion species are shaped by the charge sensitive amplifier (CSA), and their contributions overlap in the resulting waveform. Consequently, the characteristic times observed in the CSA output do not necessarily correspond directly to the physical arrival times of the negative ions. Appropriate waveform-processing methods are therefore required to recover the arrival-time difference.

In this work, we investigate two approaches for reconstructing the negative-ion arrival-time difference from simulated CSA waveforms: fast Fourier transform (FFT)-based deconvolution and direct waveform fitting, which will be introduced in Section~\ref{sec2}. In Section~\ref{sec3}, the reconstructed time differences are subsequently used to estimate the z coordinates of the ionization sources through both an analytical constant drift velocity approximation and a numerical calibration based on the simulated electric field. The primary objective is to evaluate the agreement between the reconstructed quantities and Monte Carlo truth, including reconstruction accuracy, acceptance, and the dependence of residuals on the true source position. Section~\ref{sec4} presents the reconstruction results and their comparison with Monte Carlo truth. Section~\ref{sec5} summarizes the whole paper.

\section{Signal simulation and waveform reconstruction}
\label{sec2}

\subsection{Simulated waveforms}

In our previous work~\cite{Liang:2025vkv}, an FFT-based fitting method was developed to determine the parameters of an effective readout response from Topmetal-S output waveforms. The obtained response function was then used to model the shaping of the induced current generated by drifting negative ions.

The induced currents of SeF$_5^-$ and SeF$_6^-$ are simulated using Garfield++~\cite{schindler2018garfield++} with electric and weighting fields imported from COMSOL~\cite{COMSOL}. The charge is partitioned between the two species in an assumed ratio of 1:4. The total induced current is expressed as

\begin{equation}
    I_{\mathrm{total}}(t)=I_{\mathrm{SeF}_5^-}(t)+I_{\mathrm{SeF}_6^-}(t)
\end{equation}

The corresponding CSA output waveform is obtained by convolving the total induced current with the effective response:

\begin{equation}
    V_{\mathrm{CSA}}(t)=\bigl(I_{\mathrm{total}}*f_{\mathrm{eff}}\bigr)(t).
\end{equation}

\begin{equation}
    f_{\mathrm{eff}}(t)=Ag\,h_{\tau,n}(t), \qquad
    h_{\tau,n}(t)=e^n\left(\frac{t}{n\tau}\right)^n e^{-t/\tau}
    \quad (t>0),
    \label{eq:csa-effective-response}
\end{equation}

with $h_{\tau,n}(t)=0$ for $t\leq0$. The shape has a decay time constant $\tau$, a dimensionless exponent $n$, and a continuous maximum at $t=n\tau$. In the historical Topmetal-S pulse fit, the convolved template was divided by its positive sampled maximum before applying the pulse amplitude $A$ and the time shift $t_0$:

\begin{equation}
    V_{\mathrm{fit}}(t)=A\,\frac{C(t-t_0)}{\max_k C(t_k)},
\end{equation}
where $C$ is the discrete convolution of the stored input template with $g h_{\tau,n}$, evaluated by interpolation at $t-t_0$. For the fitted polarity, the magnitude of $g$ cancels between the numerator and denominator. The fits thus provide a pulse amplitude and shape parameters. An absolute charge-to-voltage gain would require a separate calibration.

\begin{figure}
    \centering
    \includegraphics[width=0.9\linewidth]{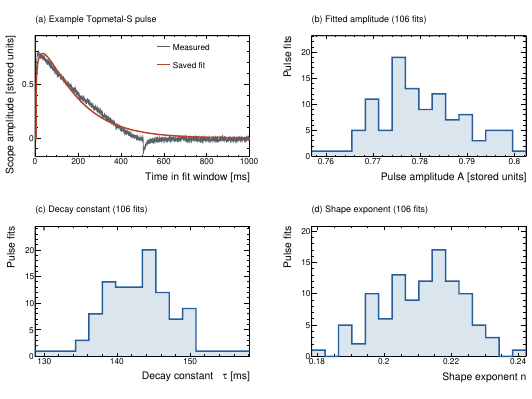}
    \caption{(a) Example Topmetal-S test pulse and its FFT-based convolution fit. The horizontal axis follows the historical fit-window convention. The measured trace is averaged in blocks of 32 samples for display. (b)--(d) Distributions of the fitted amplitude $A$, decay constant $\tau$ and shape exponent $n$ for the 106 retained fits.}
    \label{fig:Fitresult}
\end{figure}

The historical analysis assigned a nominal duration of 1~s to each fit window. The archived waveform files do not independently establish the oscilloscope sampling interval, so the time axis in the measured-pulse fit follows that convention. Of 108 fits, 106 passed the waveform-quality check. In the other two, the fitted signal was already substantial before the measured rising edge. Figure~\ref{fig:Fitresult} shows one retained fit and the distributions of $A$, $\tau$ and $n$.

The fixed CSA model used in the simulation has $\tau=143.0895$~ms, $n=0.2128095$, $A=0.7782005$, and $g=-1.72905$. The signed coefficient $g$ sets the nominal scale and polarity, and the CSA output is expressed in arbitrary response units. The discrete implementation uses zero-padded linear convolution, inverse FFT normalization, and multiplication by the simulation sampling interval. The simulated waveforms have 20,000 samples at 0.5~ms intervals, spanning 10~s.

For the response-mismatch study, one joint triplet $(A_j,\tau_j,n_j)$ was drawn uniformly with replacement from the 106 retained fits per pad and used to shape both ion currents. Drawing the triplet together preserves the fitted parameter correlations. Let $p(\tau,n)$ be the maximum of $h_{\tau,n}$ sampled at the simulation interval. The amplitude used for shaping is $A_j p(\tau_{\mathrm{nom}},n_{\mathrm{nom}})/p(\tau_j,n_j)$, where the subscript $\mathrm{nom}$ denotes the fixed CSA model. This corrects the relative pulse height for the finite sampling of the kernel. The signed coefficient $g$ is held fixed.

\begin{figure}
    \centering
    \includegraphics[width=\linewidth]{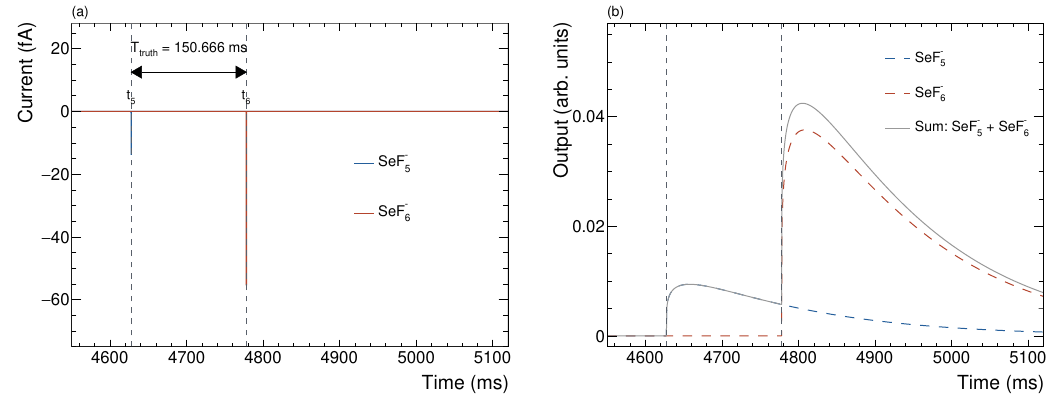}
    \caption{(a) Simulated SeF$_5^-$ and SeF$_6^-$ induced currents for the example deposition. The dashed lines mark the Monte Carlo arrival times. (b) The CSA components and their sum, shaped with the fitted response in Figure~\ref{fig:Fitresult}(a).}
    \label{fig:Waveform}
\end{figure}

Figure~\ref{fig:Waveform} shows the induced currents and the CSA waveforms for a single deposition at $z=90.5$~cm, shaped with the response from the test-pulse fit in Figure~\ref{fig:Fitresult}(a). The FFT--Direct benchmark in section~\ref{sec4} uses the fixed CSA response for waveform generation.

\subsection{FFT-Based Waveform Deconvolution}
\label{sec:fftdeconv}

The FFT-based de-convolution method was applied to recover the induced-current features from the simulated CSA output waveforms. The procedure consisted of waveform pre-processing, current recovery, peak identification, and arrival-time extraction.

First, the baseline of each waveform was estimated using the median of the first 200 sampling points and subtracted from the recorded signal. The waveform was also checked for activity in the baseline region and for significant residual signals near the end of the acquisition window.

\begin{figure}
    \centering
    \includegraphics[width=0.55\linewidth]{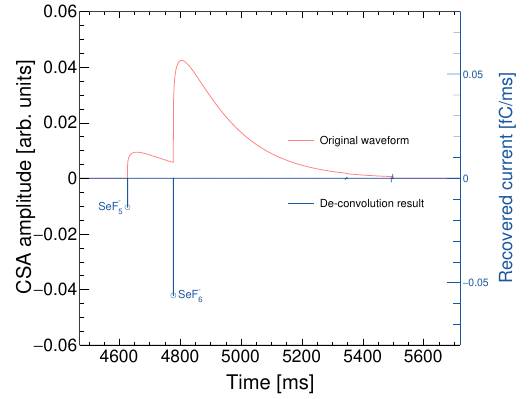}
    \caption{Summed CSA waveform and signed current recovered by FFT deconvolution for the example in Figure~\ref{fig:Waveform}. The sampled-response waveform was deconvolved with the nominal response. Markers identify the two negative-current peaks.}
    \label{fig:fft_dec}
\end{figure}

The baseline-subtracted waveform was deconvolved using the nominal CSA response in Eq.~\eqref{eq:csa-effective-response}, including when the waveform had been generated with a sampled response. A regularized FFT inversion with $\rho=10^{-12}$ was employed. Zero-padding was applied to avoid circular-convolution artifacts, and the recovered current was retained over the original 20,000 samples. Figure~\ref{fig:fft_dec} shows the recovered current features for the example in Figure~\ref{fig:Waveform}.

\begin{figure}
    \centering
    \includegraphics[width=\linewidth]{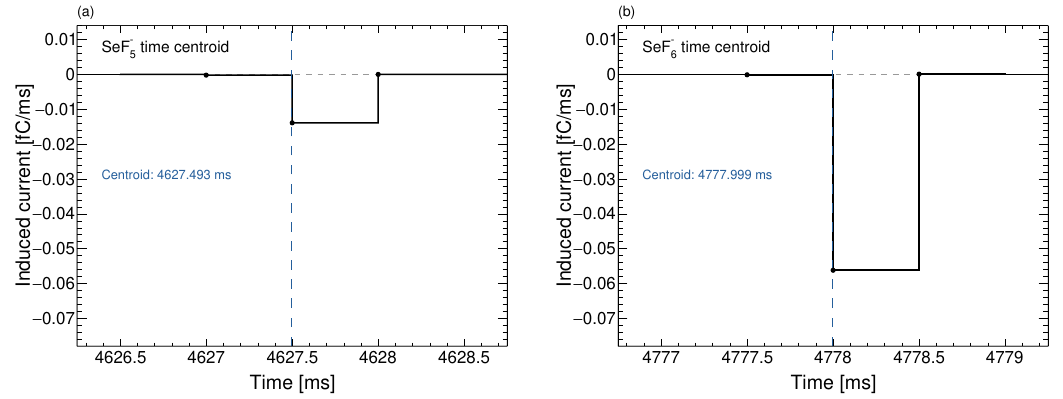}
    \caption{Three-sample centroids for (a) SeF$_5^-$ and (b) SeF$_6^-$. The markers show the samples entering each centroid and the dashed lines indicate the uncorrected centroid times. Their separation is 150.506~ms. Applying the species-dependent offsets gives a reconstructed arrival-time separation of 150.624~ms.}
    \label{fig:TimeSearch}
\end{figure}

Candidate peaks were required to exceed 5\% of the dominant negative-current peak in both height and prominence, with a minimum peak-index separation of three samples (1.5~ms). The earlier and later features were assigned to SeF$_5^-$ and SeF$_6^-$, respectively. The time $c_i$ of each feature is the charge-weighted centroid of the peak sample and its two neighbors, as shown in Figure~\ref{fig:TimeSearch}. The reconstructed arrival time is $t_i^{\mathrm{rec}}=c_i+b_i$, with the independently calibrated offsets $b_5=0.239558$~ms and $b_6=0.357868$~ms. For this example, the uncorrected centroid separation is 150.506~ms and the corrected separation $\Delta t_{\mathrm{rec}}$ is 150.624~ms. The corresponding Monte Carlo separation is 150.666~ms.

A reconstructed arrival time difference was accepted only when two valid current features were identified and the waveform quality criteria were satisfied. Waveforms failing these requirements were retained as unresolved cases. The accepted time differences were subsequently used for z coordination reconstruction.

\subsection{Direct CSA Waveform Fitting}
\label{sec:directfit}

In addition to FFT-based deconvolution, a direct waveform-fitting method was developed to extract the arrival-time difference between SeF$_5^-$ and SeF$_6^-$ from the simulated CSA output. Instead of recovering the induced current, this method fits the observed waveform with a combination of two independently simulated ion-signal templates.

A library of 12 independent simulated events was prepared to construct the waveform templates. For each event, the individual CSA responses of SeF$_5^-$ and SeF$_6^-$ were stored separately. Each template was normalized to its positive peak amplitude, and its time axis was referenced to the corresponding simulated ion-arrival time. The target event itself was not used to construct its fitting templates.

The total CSA waveform was modeled as

\begin{equation}
    V_{\mathrm{fit}}(t)
    =
    A_5 T_5(t-t_5)
    + A_6 T_6(t-t_6)
    + B,
    \label{eq:direct_fit}
\end{equation}

where $T_5$ and $T_6$ are the normalized CSA waveform templates for SeF$_5^-$ and SeF$_6^-$, respectively. The parameters $A_5$ and $A_6$ adjust the amplitudes of the two components, $t_5$ and $t_6$ determine their time positions, and $B$ represents a constant baseline.

The fitting interval was selected from the observed waveform by identifying the region exceeding $10^{-5}$ of its peak amplitude and extending the interval by 1000~ms on both sides, subject to the recorded time boundaries. The original waveform was not resampled.

For each target event, all 12 template pairs were tested. To reduce the dependence on the initial parameter values, 11 initial configurations were used for each template pair, giving 132 optimization attempts per event. The initial values were determined from the observed waveform features and the corresponding template shapes, without using the target-event arrival-time truth.

The fitting parameters were optimized using the MIGRAD algorithm, followed by HESSE for covariance evaluation~\cite{James:1975dr}. Candidate solutions were evaluated according to convergence, covariance quality, parameter boundaries, and the physical ordering $t_5<t_6$. Among eligible solutions, the result with the smallest scaled sum of squared residuals was selected. Competing solutions with substantially different arrival-time differences and comparable fitting scores were flagged as ambiguous.

The reconstructed arrival-time difference was calculated as

\begin{equation}
    \Delta t_{\mathrm{rec}}
    =
    t_6^{\mathrm{fit}}-t_5^{\mathrm{fit}}.
    \label{eq:direct_deltat}
\end{equation}

\begin{figure}
    \centering
    \includegraphics[width=0.3\linewidth]{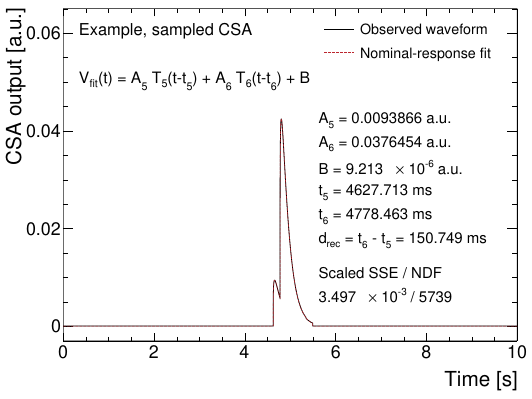}
    \includegraphics[width=0.3\linewidth]{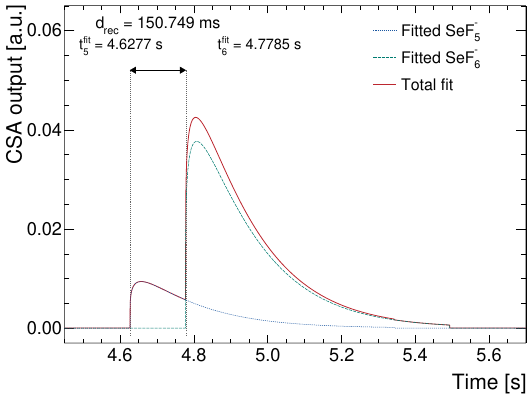}
    \includegraphics[width=0.3\linewidth]{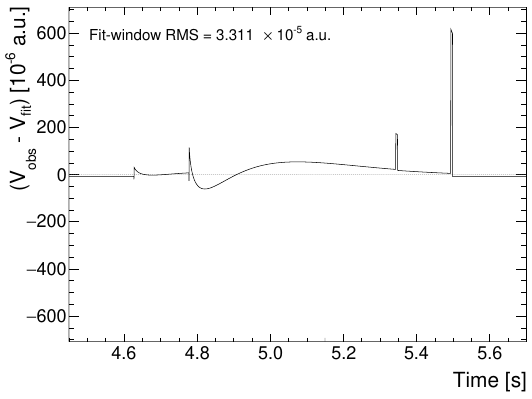}
    \caption{Direct fit to the example waveform shown in Figure~\ref{fig:fft_dec}. The fit uses the nominal CSA templates. The panels show the full 10~s observation and fit, the fitted SeF$_5^-$ and SeF$_6^-$ contributions and the total fit, and the residual $V_{\mathrm{obs}}-V_{\mathrm{fit}}$. The fitted separation is 150.749~ms.}
    \label{fig:directfit}
\end{figure}

Figure~\ref{fig:directfit} shows the direct fit to the same sampled-response waveform used in the FFT example. The fit uses the nominal CSA templates. Its fitted arrival-time separation is 150.749~ms.

\section{z coordinate reconstruction}
\label{sec3}

\subsection{Electric Field and Ion Drift}
\label{subsec:field_drift}

\begin{figure}
    \centering
    \includegraphics[width=0.5\linewidth]
    {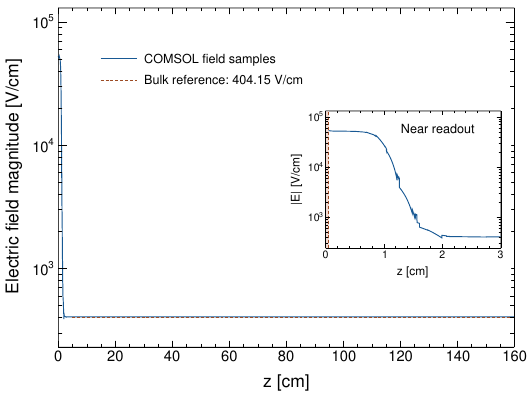}
    \caption{Electric field along the z axis at $(x,y)=(0,0)$, obtained from the COMSOL field map. The inset shows the near-readout region, where the electric field deviates substantially from the nominal value of $404.155$~V/cm.
    }
    \label{fig:ElectricField}
\end{figure}

The reduced mobilities used for SeF$_5^-$ and SeF$_6^-$ are $K_{0,5}=0.444$ and $K_{0,6}=0.430~\mathrm{cm^2\,V^{-1}\,s^{-1}}$~\cite{Liang:2025vkv}, referenced to $T_0=273.15$~K and $P_0=760$~Torr. The production simulation uses $T=293.15$~K and $P=7600$~Torr ($10$~atm, or $10.1325$~bar). In the ideal-gas density convention, the mobilities at the simulation conditions are
\[
    K_i=K_{0,i}\frac{T}{T_0}\frac{P_0}{P},
\]
giving $K_5=0.047650961$ and $K_6=0.046148453~\mathrm{cm^2\,V^{-1}\,s^{-1}}$. The field map in Figure~\ref{fig:ElectricField} has a nominal bulk magnitude of $404.155$~V/cm. The field increases near the readout plane due to the focusing structure.

\begin{figure}
    \centering
    \includegraphics[width=0.5\linewidth]
    {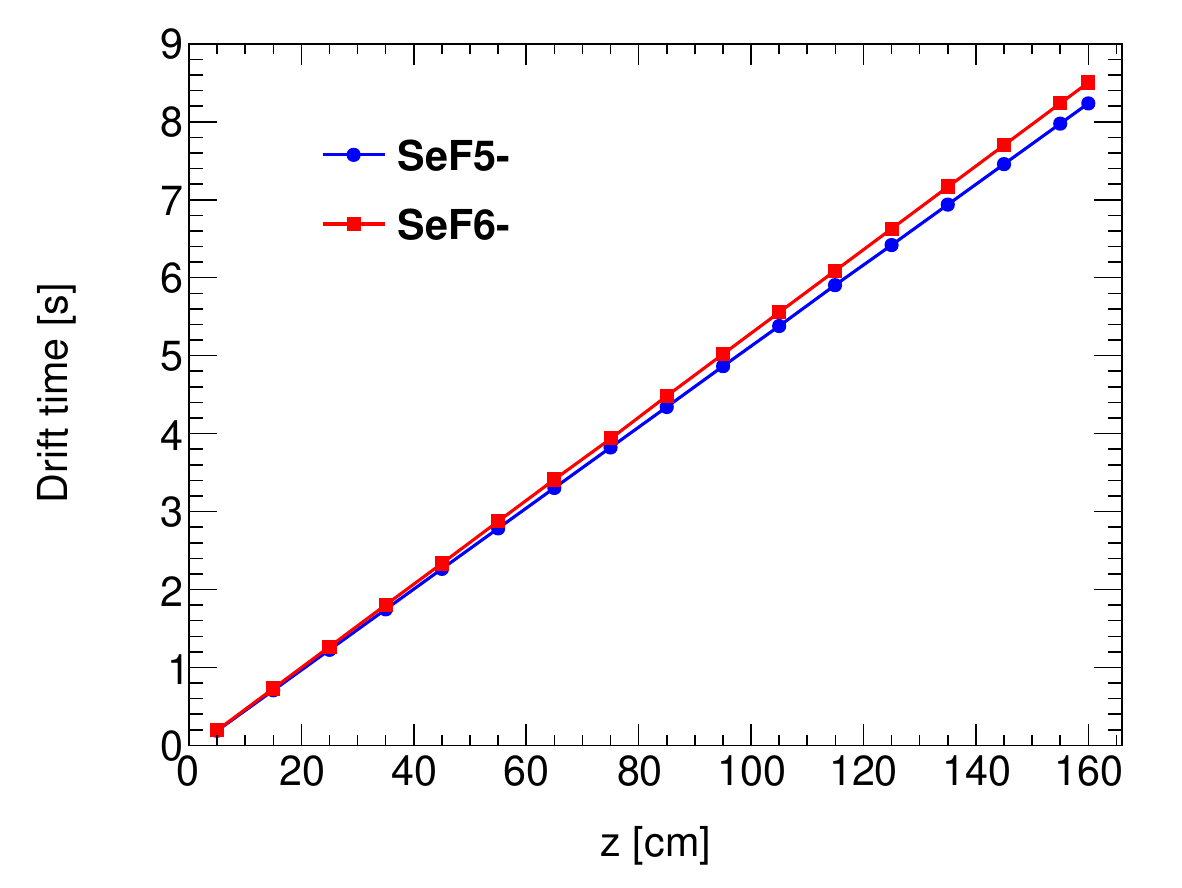}
    \caption{
    Mean drift times of SeF$_5^-$ and SeF$_6^-$ as functions of the initial z coordinate, obtained from Garfield++ simulations. At each of the 17 sampled $z$ positions, the mean drift time is calculated from 24 transverse starting positions satisfying the common readout-pad selection criterion. Error bars represent one standard deviation of the drift times across these positions.
    }
    \label{fig:DriftTime}
\end{figure}

Figure~\ref{fig:DriftTime} shows the simulated drift times as functions of the initial $z$ coordinate. At each of the 17 heights, the figure uses 24 transverse starting positions that satisfy the common-pad criterion. Both species exhibit an approximately linear increase in drift time with increasing $z$.

\subsection{z coordinate Reconstruction}
\label{subsec:z_reconstruction}

In the approximately uniform bulk field, the drift-speed magnitude of each ion species is

\begin{equation}
    v_i=K_i E_{\mathrm{bulk}},
    \qquad i=5,6,
    \label{eq:drift_velocity}
\end{equation}

where $K_i$ is the mobility at the simulation conditions and $E_{\mathrm{bulk}}$ is the nominal bulk electric-field magnitude.

Using the mobilities adopted in this simulation, the bulk drift velocities are

\begin{equation}
    \begin{aligned}
        v_5 &= 19.2584~\mathrm{cm/s},\\
        v_6 &= 18.6511~\mathrm{cm/s}.
    \end{aligned}
    \label{eq:ion_velocities}
\end{equation}

Under the constant-velocity approximation, the arrival-time difference is related to the initial z coordinate by

\begin{equation}
    \Delta t=t_6-t_5 =
    L
    \left(
        \frac{1}{v_6}-\frac{1}{v_5}
    \right),
    \label{eq:delta_t_z}
\end{equation}

where $L=z-z_0$ and $z_0=0.051$~cm is the readout reference in the absolute Garfield++ coordinate system of section~\ref{sec:intro}. With $L$ in cm and $v_i$ in cm/s, $\Delta t$ in Eq.~\eqref{eq:delta_t_z} is in seconds.

The z coordinate can therefore be estimated as

\begin{equation}
    z_{\mathrm{A}} =
    z_0+
    \frac{\Delta t_{\mathrm{rec}}}
    {1/v_6-1/v_5}.
    \label{eq:z_constant_velocity}
\end{equation}

Here, $\Delta t_{\mathrm{rec}}$ is the arrival-time difference reconstructed from the CSA waveform. Using the drift velocities in Eq.~\eqref{eq:ion_velocities} gives

\begin{equation}
    z_{\mathrm{A}}~[\mathrm{cm}]
    =
    0.051+
    0.591507\,
    \Delta t_{\mathrm{rec}}~[\mathrm{ms}]
    \label{eq:z_analytical}
\end{equation}

This relation provides an analytical approximation for converting the measured arrival-time difference into a z coordinate.

\begin{figure}
    \centering
    \includegraphics[width=0.5\linewidth]{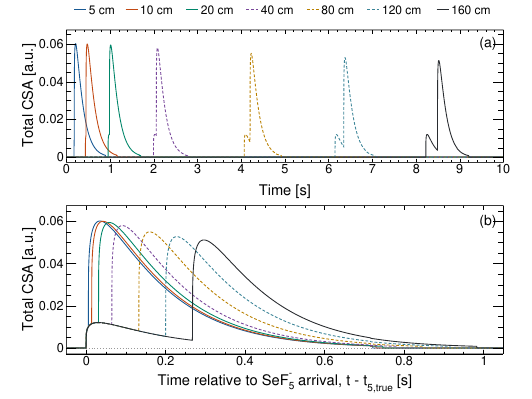}
    \caption{Waveforms in different z coordinates. Figure (a) shows the absolute time waveform from 0–10~s, while the Fig. (b) uses the arrival time of each individual SeF$_5^-$ ion as the initial point.}
    \label{fig:WavDiffz}
\end{figure}

To account for the nonuniform field near the readout, Method B was calibrated with Garfield++ transport in the COMSOL field map. The two ions in each pair shared an initial position and launch time. At each of 17 initial depths between 5 and 160~cm, 24 of the 25 transverse starting positions passed the common-pad criterion. For each accepted pair, $\Delta t$ was calculated from the simulated drift-end times output by Garfield++. These transport times were used directly, before CSA shaping and waveform reconstruction. Figure~\ref{fig:WavDiffz} shows representative waveforms generated with the fixed CSA response at different depths.

The calibration at each depth is the median physical arrival-time separation over the accepted transverse positions,

\begin{equation}
    \Delta t_{\mathrm{B}}(z)
    =
    \operatorname{median}_{(x,y)}
    \left[
        t_{6,\mathrm{arr}}(x,y,z)-t_{5,\mathrm{arr}}(x,y,z)
    \right].
    \label{eq:numerical_calibration}
\end{equation}

The calibration points were interpolated using a piecewise cubic Hermite interpolating polynomial (PCHIP). The z coordinate is then reconstructed by numerically inverting this relation:

\begin{equation}
    z_{\mathrm{B}} =
    \Delta t_{\mathrm{B}}^{-1}
    \left(\Delta t_{\mathrm{rec}}\right).
    \label{eq:z_numerical}
\end{equation}

The median curve is defined over 5--160~cm. In the performance evaluation the inverse is restricted to the common supported separation interval from 6.615 to 267.824~ms, given by the intersection of the separation ranges of the retained transverse sites.

\begin{figure}[tb]
    \centering
    \includegraphics[width=0.5\linewidth]{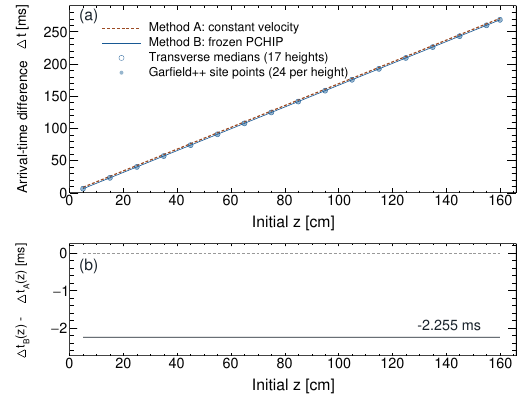}
    \caption{
    Comparison of the constant-velocity approximation (Method A) and the numerical Garfield++ calibration (Method B). Panel (a) shows the arrival-time separation versus absolute initial $z$, with individual Garfield++ drift-end separations, their medians at each depth and the PCHIP curve through those medians. Panel (b) shows $\Delta t_{\mathrm{B}}(z)-\Delta t_{\mathrm{A}}(z)$ at the same initial $z$.
    }
    \label{fig:ZCalibration}
\end{figure}

Figure~\ref{fig:ZCalibration} compares the constant-velocity approximation (Method A) with the numerical calibration (Method B). The two relations have nearly identical slopes over the simulated drift region. The numerical calibration predicts an arrival-time difference approximately $2.255$~ms smaller than the constant-velocity approximation at the same initial $z$.

For a given reconstructed arrival-time difference, Method B gives a depth about $1.334$~cm greater than Method A. This illustrates the limitations of extending the bulk drift velocities through the nonuniform readout region. Identifying the physical origin of the offset would require a separate analysis of that region.

In the following analysis, the numerical calibration is adopted as the primary mapping from the reconstructed arrival-time difference to the z coordinate. The analytical relation is retained as a reference for evaluating the influence of the constant-velocity approximation.

Let $d_{\Delta t}$ denote a paired change in reconstructed separation between sampled and fixed CSA shaping, and $d_{z,\mathrm{A}}$ and $d_{z,\mathrm{B}}$ the corresponding changes in mapped depth. Then
\[
    d_{z,\mathrm{A}}\,[\mathrm{cm}]
      =0.591507\,d_{\Delta t}\,[\mathrm{ms}], \qquad
    d_{z,\mathrm{B}}
      \simeq
      \left[\left.\frac{\mathrm{d}\Delta t_{\mathrm{B}}}{\mathrm{d}z}
      \right|_{z_{\mathrm{B}}^{\mathrm{fixed}}}\right]^{-1}d_{\Delta t}.
\]
The first relation follows from the linear calibration. The second is the local approximation for the inverse PCHIP relation, with both reconstructions restricted to the supported range. Under both shaping conditions, FFT uses the same nominal inversion kernel and direct fitting uses the same nominal templates. The depth mappings are also held fixed. Their paired changes therefore measure the incremental effect of changing the shaping response for the same simulated depositions.

\section{Reconstruction results}
\label{sec4}

In this section, the reconstruction performance is evaluated by comparing the reconstructed quantities with Monte Carlo truth. We first examine the arrival-time difference and z coordinate for single energy depositions, followed by the reconstruction of complete three-dimensional tracks.

\subsection{Comparison of arriving time and z coordinate}

The arrival-time reconstruction methods introduced in section~\ref{sec2} were evaluated using simulated single-deposition waveforms generated with the fixed nominal CSA response. A total of 873 waveforms satisfied the geometric selection criteria and were processed using both FFT deconvolution and direct waveform fitting. Among these waveforms, 855 were accepted by the FFT method, while all 873 were accepted by direct fitting.

\begin{figure}
    \centering
    \includegraphics[width=1.\linewidth]{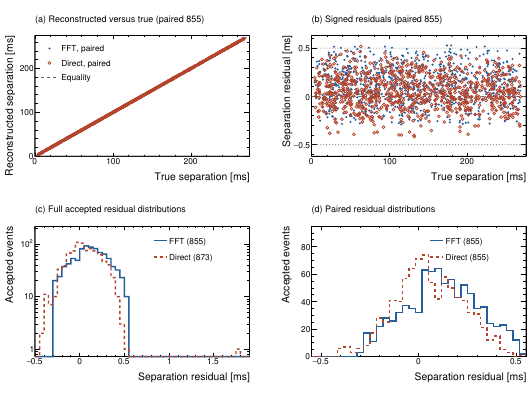}
    \caption{Comparison of reconstructed and simulated arrival-time differences for single depositions generated with the fixed nominal CSA response. (a) Reconstructed versus true arrival-time separation for the 855 events accepted by both methods. The dashed line indicates equality. (b) Signed reconstruction residual versus true separation for the same events. (c) Full residual distributions for FFT deconvolution (855 events) and direct waveform fitting (873 events), shown on a logarithmic count scale. (d) Residual distributions for the common set of 855 accepted events. Blue and orange represent FFT deconvolution and direct waveform fitting, respectively. The full Direct distribution in (c) includes the large positive residual of +1.79 ms.}
    \label{fig:Timeing}
\end{figure}

Figure~\ref{fig:Timeing} compares the reconstructed arrival-time differences with the corresponding Monte Carlo truth. The reconstruction error is defined as $\delta(\Delta t)=\Delta t_{\mathrm{reco}}-\Delta t_{\mathrm{truth}}$. For a consistent comparison, the 855 waveforms accepted by both methods were used to evaluate their timing accuracy. The mean reconstruction errors are $+0.118$~ms for FFT de-convolution and $+0.052$~ms for direct fitting, with corresponding standard deviations of $0.188$ and $0.172$~ms. The root-mean-square errors are $0.222$ and $0.180$~ms, respectively. Direct waveform fitting therefore provides a smaller reconstruction error for the common sample.

The residual distributions in Figure~\ref{fig:Timeing}(c,d) show the differences between the two methods more clearly. Using an absolute-error criterion of $0.5$~ms, 845 FFT results and 854 direct-fitting results satisfy the criterion within the common sample. However, an accepted fit does not necessarily guarantee an accurate arrival-time difference. In the complete set of 873 direct-fitting results, one event exhibits a residual of $+1.79$ ms, as shown in Figure~\ref{fig:Timeing}(c). This illustrates the importance of evaluating the reconstruction accuracy against the simulated arrival times rather than relying only on the fitting acceptance criteria.

\begin{figure}
    \centering
    \includegraphics[width=1.\linewidth]{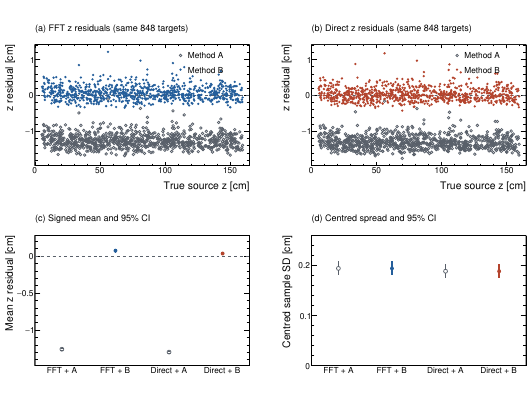}
    \caption{Depth reconstruction for single depositions generated with the fixed nominal CSA response. (a) FFT and (b) Direct signed depth residuals versus the simulated source depth. Open grey circles represent the constant-velocity calibration (Method A), while filled symbols represent the numerical calibration (Method B). (c) Mean signed depth residuals and (d) centred sample standard deviations for the four reconstruction combinations, with 95\% paired-bootstrap confidence intervals. All combinations are evaluated using the same 848 single-deposition events.}
    \label{fig:depth}
\end{figure}

The reconstructed arrival-time differences were subsequently converted into $z$ coordinates using Methods A and B described in section~\ref{sec3}. The common sample of 848 depositions consists of events accepted by both timing estimators for which both depth mappings give valid reconstructed positions. Figure~\ref{fig:depth} shows the $z$ reconstruction residuals, defined as $\delta z=z_{\mathrm{reco}}-z_{\mathrm{truth}}$, as functions of the true source position.

Using the constant velocity approximation (Method A), the mean z residuals are $-1.259$ cm for FFT de-convolution and $-1.297$~cm for direct fitting, with corresponding RMSEs of $1.273$ and $1.311$~cm. After applying the numerical calibration (Method B), the mean residuals decrease in magnitude to $+0.075$ and $+0.036$~cm, respectively. The corresponding RMSEs are reduced to $0.208$ and $0.192$~cm.

As shown in Figure~\ref{fig:depth}(c), Method B substantially reduces the systematic bias in the reconstructed z position. However, the central standard deviations remain approximately $0.194$~cm for FFT and $0.189$~cm for direct fitting with either calibration method, as shown in Figure~\ref{fig:depth}(d). This indicates that the improvement in RMSE originates mainly from the correction of the systematic position offset, rather than a reduction in the spread of the reconstruction errors. 

The results show that the arrival-time difference can be reconstructed with sub-millisecond errors for the majority of the accepted simulated single-deposition waveforms. The numerical depth calibration further reduces the systematic error in the z position. These results characterize the reconstruction accuracy under the assumed signal and detector models, rather than the experimentally achievable timing or spatial resolution.

\begin{figure}[tb]
    \centering
    \includegraphics[width=\linewidth]{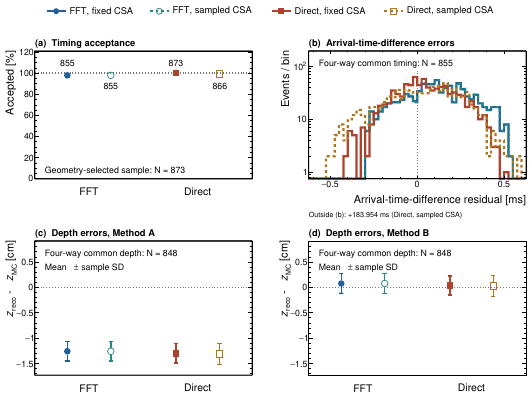}
    \caption{Single-deposition reconstruction with fixed and sampled CSA shaping. (a) Timing acceptance among 873 geometry-selected inputs. (b) Separation residuals on the four-way common timing sample of 855 events. (c,d) The mean and centred sample standard deviation of the depth residuals under Methods A and B on the 848 events valid for all reconstructions and both mappings. The near-readout outlier lies outside the common samples because FFT rejected it at the pretrigger check. Its accepted sampled-response direct-fit residual is included in the full-sample statistics.}
    \label{fig:csa-four-way}
\end{figure}

\paragraph{Sensitivity to the shaping response.}
For the same 873 geometry-selected depositions, the stored currents were shaped with one draw per event from the measured-pulse library (seed 1043). Both estimators processed the same sampled waveforms. FFT used the nominal kernel and direct fitting used the nominal template library. FFT accepted the same 855 events (97.94\%) under both shaping conditions. The other 18 failed the pretrigger quiet-window requirement. Direct fitting retained 866 (99.20\%) with sampled shaping and 873 with fixed shaping. The nonacceptances comprised 7 ambiguous fits. These events also failed the FFT pretrigger check. Table~\ref{tab:csa-response-mismatch} and Figure~\ref{fig:csa-four-way} compare the four configurations. The common samples contain 855 timing targets and 848 depth targets. Method B primarily reduces the mean depth bias for both shaping conditions.

One near-readout deposition illustrates the difference between the common and full acceptance samples. Its simulated separation is 0.323~ms. The direct fit to the sampled waveform returned 184.277~ms and passed the original quality selection. FFT rejected the waveform at the pretrigger check. The event contributes to the full sampled-response direct-fit timing RMSE of 6.255~ms and Method B depth RMSE of 3.737~cm. The corresponding RMSEs on the common samples are 0.217~ms and 0.206~cm.

Define the paired shift $d_i=t_{i,\mathrm{sampled}}^{\mathrm{rec}}-t_{i,\mathrm{fixed}}^{\mathrm{rec}}$. For FFT, the RMS shifts of $t_5$, $t_6$, and their separation are 2.766, 2.788, and 0.460~$\mu$s. Method A gives an FFT depth-shift RMS of 2.720~$\mu$m, consistent with the linear propagation in section~\ref{sec3}. Method B gives 2.724~$\mu$m on 848 FFT pairs. For direct fitting, the three timing shifts have RMS values of 0.166, 6.250, and 6.243~ms on 866 paired events. The depth-shift RMS is 3.693~cm for Method A and 0.111~cm for Method B. The latter uses 848 pairs and excludes this near-readout outlier, whose fixed-response estimate is below the inverse domain. These shifts measure the effect of changing the shaping response for the same deposition.

\input{FourWayPerformanceTable.tex}

\subsection{3D track reconstruction}

For a complete Geant4 event, multiple energy depositions at different positions may contribute to the signal recorded by the same readout pad. The resulting CSA waveform contains overlapping contributions from multiple ionization sites. The full-track waveforms and the two-ion fit templates use the fixed nominal CSA response.

\begin{figure}
    \centering
    \includegraphics[width=0.5\linewidth]{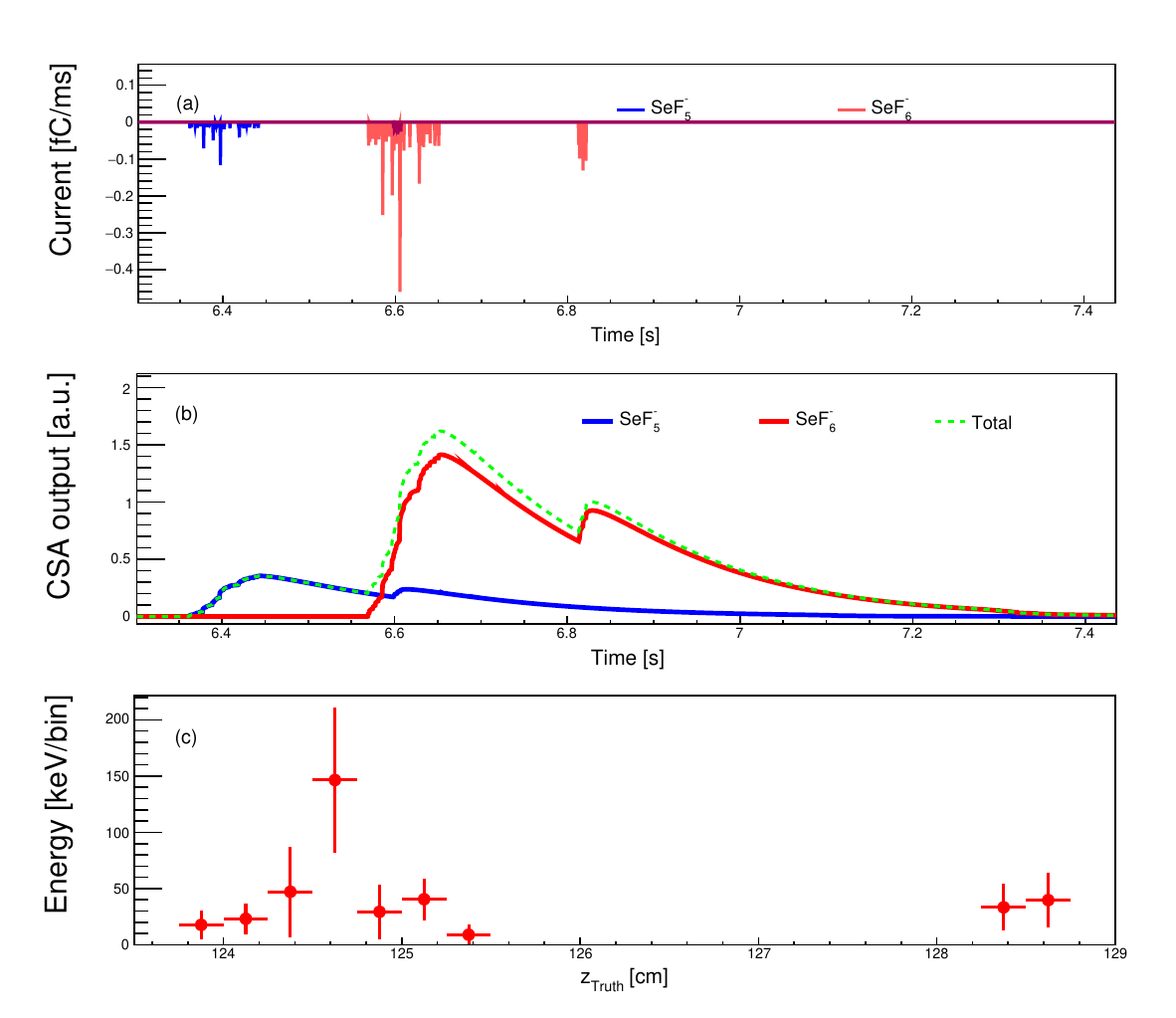}
    \caption{An example of multiple energy depositions contributing to a single readout pad, which includes the induced current and corresponding CSA response, as well as the energy deposition. }
    \label{fig:PadSig}
\end{figure}

Figure~\ref{fig:PadSig} shows the induced currents and the individual and summed CSA responses for a representative pad, together with the distribution of deposited energy in the true source depth. Deposits at several depths contribute to the same pad waveform. Consequently, the arrival-time information cannot generally be represented by one pair of ion-arrival times. The reconstruction therefore aims to recover an
effective z coordinate from the complete CSA waveform.

The z coordinate is obtained by fitting the full CSA waveform
on each pad. For pad \(p\) in event \(e\), the fitted waveform and effective depth are

\begin{equation}
\begin{aligned}
\widehat V_{ep}(t) &= B_{ep}
 + \sum_k q_{epk}\,S(z_{epk},t-t_{0,e}),\\
z_{\mathrm{eff},ep} &= \frac{\sum_k q_{epk}z_{epk}}{\sum_k q_{epk}},
 \qquad q_{epk}\geq 0 .
\end{aligned}
\label{eq:full-track-effective-depth}
\end{equation}

Here \(S\) is the two-ion CSA template for a 10 keV deposition at depth \(z_{epk}\), \(q_{epk}\) is the fitted amplitude in template units, and \(B_{ep}\) is the baseline. The templates cover \(z=5\)--160 cm. The z coordinate distribution is represented by a single depth, a broadened distribution or two discrete depths. The time shift \(t_{0,e}\) is fitted jointly for all pads in the event. The transverse coordinates are assigned to the pad centers.

The shared event time was evaluated using the mean pad loss. For each pad, the loss was defined as the full-record sum of squared residuals divided by the squared pulse peak and the number of samples in the signal region. This region extends from the first to the last sample above 0.1\% of the pulse peak, with a 50~ms margin on each side.

A single-depth model was retained if it met the waveform-fit criteria. Otherwise, a broadened model could replace it if the loss decreased by at least 20\%. If the broadened model failed the waveform-fit criteria, a two-depth candidate was tested. It was selected only if it reduced the loss by at least 20\% relative to the current candidate, the two depths were separated by at least 1~cm, and each carried at least 15\% of the total fitted weight.

A fit is retained if it converges with positive total amplitude, its signal-region RMS and maximum absolute residual are at most 4\% and 20\% of the pulse peak, and the depth range across the tested alternatives is at most 2~cm. Fits at the boundary of the common-time search are excluded. The Monte Carlo reference depth \(z_{\mathrm{MC},ep}\) is the energy-weighted mean source depth of the deposits for which both species were routed to that pad.

Of 105 consecutive Geant4 events screened, 100 satisfy the source
selection. All positive-energy deposits in these events lie in the drift gas with \(5\leq z\leq160\) cm, \(|x|,|y|<50\) cm and \(0\leq t<10\) s. Each event spans at least two transverse \(1\,\mathrm{cm}\times 1\,\mathrm{cm}\) cells, with multiple deposits in at least one cell. The fit retains 1,185 of 1,810 observed pads (65.47\%). Deposits routed to these pads by both species account for 66.49\% of the total deposited energy of the selected events.

\begin{figure}
    \centering
    \includegraphics[width=\linewidth]{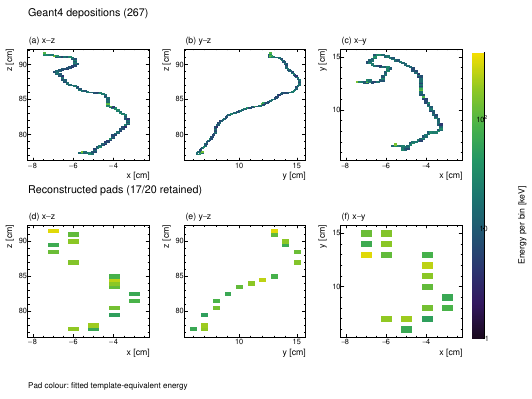}
    \caption{Track projections for an example Geant4 event in the (x)–(z), (y)–(z) and (x)–(y) planes. Panels (a)–(c) show the energy deposited by all 267 Geant4 steps in (0.125~cm $\times$ 0.125~cm) bins. Panels (d)–(f) show the 17 retained pad positions in (0.5~cm $\times$ 0.5~cm) bins, weighted by the fitted template-equivalent energy. All panels share a logarithmic energy scale in keV per bin.}
    \label{fig:full-track-example}
\end{figure}

Figure~\ref{fig:full-track-example} shows an example of the truth and reconstructed track projections. Of the 20 observed pads, 17 pass the fit-quality criteria. Their z positions have an RMSE of 0.090 cm relative to the corresponding energy-weighted Geant4 pad depths.

%The effective arrival times are calculated from the fitted z coordinate distribution and the common time shift. Relative to the energy-weighted simulated endpoint times, the pad-level RMSEs for \(\mathrm{SeF}_5^-\), \(\mathrm{SeF}_6^-\) and their time difference are 2.811, 2.138 and 2.165 ms, respectively. The RMSE of \(t_{0,e}\) is 72.920 ms relative to the zero additional time shift in the simulated waveforms. The common shift cancels explicitly in the time difference, while the fitted z distribution still depends on it.

\begin{figure}
    \centering
    \includegraphics[width=\linewidth]{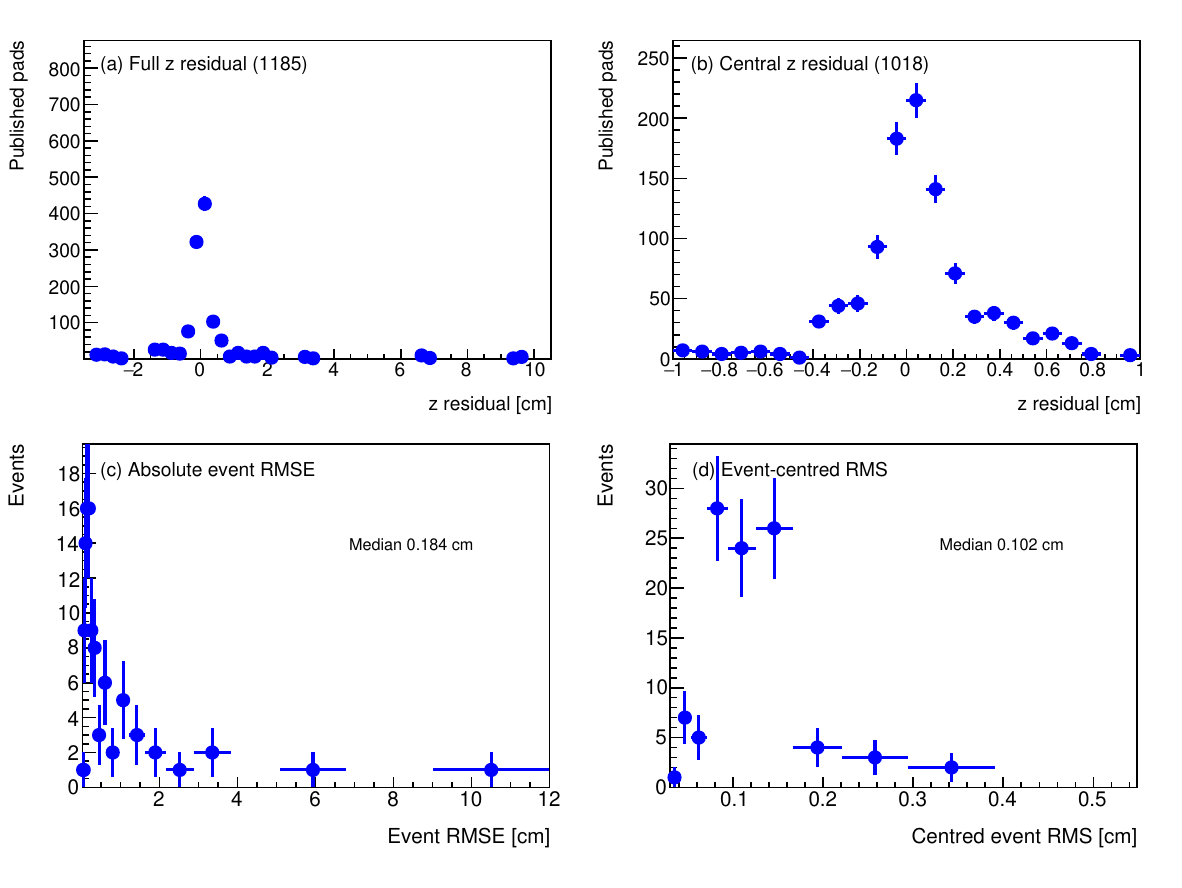}
    \caption{Reconstructed z coordinate residuals for 100 events. (a) Full distribution for the 1,185 retained pads and (b) the central interval \(\lvert r_{ep}\rvert\leq1\) cm. Panels (c) and (d) show the event RMSE before and after subtracting each event's mean residual.}
    \label{fig:full-track-depth-errors}
\end{figure}

Figure~\ref{fig:full-track-depth-errors} shows the reconstructed z residual \(r_{ep}=z_{\mathrm{eff},ep}-z_{\mathrm{MC},ep}\). The median event RMSE is 0.184~cm, and the RMSE over all retained pads is 1.287~cm. The signed residuals range from \(-3.195\) to \(+9.731\) cm. Subtracting the mean residual of each event reduces the median event RMS to 0.102~cm. This quantity describes the longitudinal track shape independently of the mean offset.

\section{Discussion and Conclusion}
\label{sec5}

For single depositions shaped with the fixed response, direct waveform fitting gives a smaller timing RMSE on the sample accepted by both estimators. The numerical depth calibration mainly changes the intercept of the time-to-depth relation over the common calibration range. This reduces the mean depth bias and the RMSE while leaving the centred residual width nearly unchanged.

\begin{figure}[tbp]
    \centering
    \includegraphics[width=\linewidth]{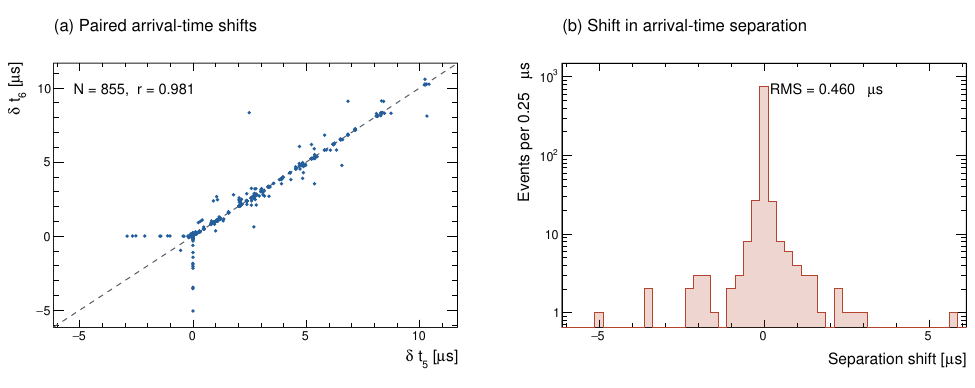}
    \caption{Paired changes in the FFT timing estimates, defined as sampled minus fixed CSA shaping, for the 855 events accepted in both conditions. (a) Shifts in the two arrival times. The dashed line denotes equality. (b) Distribution of the shift in their separation, on a logarithmic count scale.}
    \label{fig:fft-paired-timing-shifts}
\end{figure}

The paired FFT arrival-time shifts have a Pearson correlation of 0.981 (Figure~\ref{fig:fft-paired-timing-shifts}). A common shift cancels in $d_6-d_5$. The corresponding correlation for direct fitting is 0.079. On the four-way common timing sample, the direct-fit correlation is 0.417, with a separation-shift RMS of 187.3~$\mu$s. The near-readout outlier dominates the separation-shift RMS over the full direct-fit paired sample. The FFT Method B shift RMS of 2.724~$\mu$m measures the change between shaping conditions. The 0.208~cm RMSE measures the error relative to Monte Carlo truth. The test covers the response variation represented by the retained pulse fits under one seeded assignment, with both reconstruction procedures held fixed. Electronics noise and diffusion are absent.

For complete tracks, the main limitation is the coupling between the common event time and the absolute depth. A shift in the fitted event time can be compensated by a displacement of the reconstructed track along the drift direction, with little change in either the waveform residuals or the relative positions of the pads. This behaviour explains why some events have small centered depth errors despite large absolute offsets. The centred residuals assess the relative longitudinal shape, whereas the uncentered residuals determine how accurately the track is placed in the detector. An independent event-time constraint would help distinguish between competing absolute-depth solutions.

In summary, the simulated Topmetal-S waveforms support arrival-time-difference reconstruction for single depositions and pad-level reconstruction of relative longitudinal structure for full tracks. Using sampled shaping preserved FFT acceptance and exposed a loss of acceptance in direct fitting under the frozen selection rules. These results are conditional on the assumed ion populations, transport model and waveform selection. Measurements with detector signals are needed to determine how diffusion, electronics noise and the response of neighbouring pads affect the achievable timing and spatial resolution.

\acknowledgments

% Bibliography

%% [A] Recommended: using JHEP.bst file
%% \bibliographystyle{JHEP}
%% \bibliography{biblio.bib}

%% or
%% [B] Manual formatting (see below)
%% (i) We suggest to always provide author, title and journal data or doi:
%% in short all the informations that clearly identify a document.
%% (ii) please avoid comments such as "For a review'', "For some examples",
%% "and references therein" or move them in the text. In general, please leave only references in the bibliography and move all
%% accessory text in footnotes.
%% (iii) Also, please have only one work for each \bibitem.
\bibliographystyle{ieeetr}
\bibliography{biblio}
\end{document}

%% file: FourWayPerformanceTable.tex
\begin{table}[tb]
    \centering
    \caption{Single-deposition results for fixed and sampled CSA shaping. The first block uses each configuration's own outcomes on the same 873 inputs. The remaining rows give the signed mean, centred sample standard deviation, and RMSE relative to Monte Carlo truth on the 855 common timing targets and 848 common depth targets.}
    \label{tab:csa-response-mismatch}
    \small
    \setlength{\tabcolsep}{5pt}
    \begin{tabular}{lrrrr}
        \hline
        Quantity & FFT, fixed & Direct, fixed & FFT, sampled & Direct, sampled \\
        \hline
        Accepted times & 855 & 873 & 855 & 866 \\
        Valid Method A depths & 855 & 873 & 855 & 866 \\
        Valid Method B depths & 848 & 848 & 848 & 849 \\
        Own $\Delta t$ RMSE [ms] & 0.221752 & 0.190166 & 0.221803 & 6.254766 \\
        Own Method B RMSE [cm] & 0.208253 & 0.191980 & 0.208273 & 3.737431 \\
        \hline
        Mean $\delta(t_5)$ [ms] & -0.0236 & 0.0832 & -0.0221 & -0.0183 \\
        SD $\delta(t_5)$ [ms] & 0.1303 & 0.1229 & 0.1303 & 0.1634 \\
        RMSE $\delta(t_5)$ [ms] & 0.1323 & 0.1484 & 0.1321 & 0.1644 \\
        \hline
        Mean $\delta(t_6)$ [ms] & 0.0942 & 0.1354 & 0.0957 & 0.0141 \\
        SD $\delta(t_6)$ [ms] & 0.1302 & 0.1242 & 0.1303 & 0.2133 \\
        RMSE $\delta(t_6)$ [ms] & 0.1607 & 0.1837 & 0.1616 & 0.2136 \\
        \hline
        Mean $\delta(\Delta t)$ [ms] & 0.1178 & 0.0522 & 0.1178 & 0.0323 \\
        SD $\delta(\Delta t)$ [ms] & 0.1880 & 0.1725 & 0.1880 & 0.2145 \\
        RMSE $\delta(\Delta t)$ [ms] & 0.2218 & 0.1801 & 0.2218 & 0.2168 \\
        \hline
        Mean $\delta(z_{\rm A})$ [cm] & -1.2585 & -1.2974 & -1.2585 & -1.3093 \\
        SD $\delta(z_{\rm A})$ [cm] & 0.1943 & 0.1886 & 0.1943 & 0.2049 \\
        RMSE $\delta(z_{\rm A})$ [cm] & 1.2734 & 1.3111 & 1.2734 & 1.3253 \\
        \hline
        Mean $\delta(z_{\rm B})$ [cm] & 0.0753 & 0.0364 & 0.0753 & 0.0244 \\
        SD $\delta(z_{\rm B})$ [cm] & 0.1943 & 0.1886 & 0.1943 & 0.2049 \\
        RMSE $\delta(z_{\rm B})$ [cm] & 0.2083 & 0.1920 & 0.2083 & 0.2063 \\
        \hline
    \end{tabular}
\end{table}

%% file: biblio.bib
@article{Liang:2025vkv,
    author = "Liang, Tianyu and others",
    title = "{Development of a simulation and analysis framework for N{\ensuremath{\nu}}DEx experiment}",
    eprint = "2508.13582",
    archivePrefix = "arXiv",
    primaryClass = "physics.ins-det",
    doi = "10.1016/j.nima.2026.171439",
    journal = "Nucl. Instrum. Meth. A",
    volume = "1087",
    pages = "171439",
    year = "2026"
}

@article{nvdexcdr2023,
	author = {Cao, X.-G. and Chang, Y.-L. and Chen, K. and others},
	date = {2023/12/22},
	doi = {10.1007/s41365-023-01360-7},
	id = {Cao2023},
	isbn = {2210-3147},
	journal = {Nucl. Sci. Tech.},
	number = {1},
	pages = {3},
	title = {{N$\nu$DEx}-100 conceptual design report},
	volume = {35},
	year = {2023},
	}

@article{Nygren_2018,
doi = {10.1088/1748-0221/13/03/P03015},
year = {2018},
month = {mar},
publisher = {},
volume = {13},
number = {03},
pages = {P03015},
author = {Nygren, D. R. and Jones, B. J. P. and López-March, N. and others},
title = {Neutrinoless double beta decay with $^{82}${SeF}$_6$ and direct ion imaging},
journal = {J. Instrum.}
}

@article{MAuger_2012,
doi = {10.1088/1748-0221/7/05/P05010},
year = {2012},
month = {may},
publisher = {},
volume = {7},
number = {05},
pages = {P05010},
author = {Auger, M. and Auty, D. J. and Barbeau, P. S. and others},
title = {The {EXO}-200 detector, {Part I}: detector design and construction},
journal = {J. Instrum.}
}

@article{Adhikari_2022,
  author       = {Adhikari, G. and Al Kharusi, S. and Angelico, E. and others},
  title        = {{nEXO}: neutrinoless double beta decay search beyond $10^{28}$-year half-life sensitivity},
  journal      = {J. Phys. G: Nucl. Part. Phys.},
  year         = {2022},
  volume       = {49},
  number       = {1},
  pages        = {015104},
  doi          = {10.1088/1361-6471/ac3631}
}

@article{next2023,
  author       = {Novella, P. and Sorel, M. and Us{\'o}n, A. and others},
  title        = {Demonstration of neutrinoless double beta decay searches in gaseous xenon with {NEXT}},
  journal      = {J. High Energy Phys.},
  year         = {2023},
  volume       = {2023},
  number       = {9},
  pages        = {190},
  doi          = {10.1007/JHEP09(2023)190},
 }

@article{PandaX2017,
  author       = {Chen, X. and Fu, C.-B. and Galan, J. and others},
  title        = {{PandaX-III}: searching for neutrinoless double beta decay with high-pressure $^{136}${Xe} gas time projection chambers},
  journal      = {Sci. China Phys. Mech. Astron.},
  year         = {2017},
  volume       = {60},
  number       = {6},
  pages        = {061011},
  doi          = {10.1007/s11433-017-9028-0},
}

@article{Liang_2024,
doi = {10.1088/1748-0221/19/04/C04004},
url = {https://dx.doi.org/10.1088/1748-0221/19/04/C04004},
year = {2024},
month = {apr},
publisher = {IOP Publishing},
volume = {19},
number = {04},
pages = {C04004},
author = {Tianyu Liang and Dongliang Zhang and Hulin Wang and Chaosong Gao and Jun Liu and Xiangming Sun and Le Xiao and Feng Liu and Bihui You and Ling Liu and Yichen Yang and Kai Chen},
title = {Performance of a novel charge sensor on the ion detection for the development of a high-pressure avalancheless ion TPC},
journal = {Journal of Instrumentation},
}

@misc{schindler2018garfield++,
  title={Garfield++—simulation of ionisation based tracking detectors},
  author={Schindler, H and Veenhof, R},
  year={2018}
}

@misc{COMSOL,
  title={COMSOL {Multiphysics} v. 5.6.},
  author={COMSOL AB, Stockholm.},
  url = {www.comsol.com}
}

@article{James:1975dr,
    author       = {James, F. and Roos, M.},
    title        = {Minuit: A System for Function Minimization
                    and Analysis of the Parameter Errors and Correlations},
    journal      = {Comput. Phys. Commun.},
    volume       = {10},
    pages        = {343--367},
    year         = {1975},
    doi          = {10.1016/0010-4655(75)90039-9}
}
